# Securely Outsourcing Large Scale Eigen Value Problem To Public Cloud

Jarin Firose Moon, Shamminuj Aktar and M.M.A. Hashem
Department of Computer Science and Engineering
Khulna University of Engineering & Technology (KUET)
Khulna 9203, Bangladesh
Email: jarinfirose@gmail.com, nuj42rahman@gmail.com and mma.hashem@outlook.com

*Abstract*—Cloud computing enables clients with limited computational power to economically outsource their large scale computations to a public cloud with huge computational power. Cloud has the massive storage, computational power and software which can be used by clients for reducing their computational overhead and storage limitation. But in case of outsourcing, privacy of client's confidential data must be maintained. We have designed a protocol for outsourcing large scale Eigen value problem to a malicious cloud which provides input/output data security, result verifiability and client's efficiency. As the direct computation method to find all eigenvectors is computationally expensive for large dimensionality, we have used power iterative method for finding the largest Eigen value and the corresponding Eigen vector of a matrix. For protecting the privacy, some transformations are applied to the input matrix to get encrypted matrix which is sent to the cloud and then decrypting the result that is returned from the cloud for getting the correct solution of Eigen value problem. We have also proposed result verification mechanism for detecting robust cheating and provided theoretical analysis and experimental result that describes high-efficiency, correctness, security and robust cheating resistance of the proposed protocol.

## I. INTRODUCTION

Cloud computing provides a large number of services to client. Among them, outsourcing large scale computations is becoming very popular now-a-days. In the context of outsourcing, computationally weak clients (or devices), limited by the slow processing speed, memory, and other hardware constraints can outsource their computations and data to cloud which are equipped with massive computational resources.

Although having tremendous benefits, as customers and cloud are not in the same trusted domain, there can be many security concerns and challenges toward this outsourcing model [1]. Here, the first challenge is to protect the security of client's confidential input/output data. Customer's confidential data may contain sensitive information. For preserving data security, the client has to encrypt their confidential data before outsourcing and after outsourcing, decrypt the returned result from cloud. The second challenge is to verify the returned result whether it is correct or not as cloud may cheat and give incorrect results. Also there may be software bugs or hardware failures in cloud server which may also result in an incorrect result. The third challenge is client's efficiency. The model should provide better efficiency to client in terms of computational cost and time.

Focusing on the engineering and scientific computing problems, Eigen value problem is a basic computational problem and has a number of applications. It is used in information retrieval [2] and Google uses the eigenvector for the maximal eigenvalue of the Google matrix to determine the rank of a page for search [3]. Eigenvectors are fundamental to principal components analysis which is commonly used for dimensionality reduction in face recognition and other machine learning applications [4]. Eigen values and Eigen vectors are also used in image processing. The matrices from these applications are normally very large. When Eigen value problem deals with matrix that is very large, outsourcing the problem to a powerful cloud is an economical solution. Again, for clients as battery-limited mobile phones, portable devices, or embedded smart cards, secure outsourcing is preferred even if the data is in a moderate scale [5].

Consequently, we are directed to design a protocol which outsources Eigen value problem to a public cloud and provides efficiency, security, robust cheating detection to the client.

The direct computation method for finding all eigenvectors takes $o(n^3)$ in time complexity, which becomes very expensive for large dimension. So, in this paper, we have formulated the problem of securely outsourcing large-scale Eigen value problem via iterative method, which is Power method [6], [7] that simultaneously fulfills the goals of high efficiency, correctness, security, robust cheating detection [5]. Our mechanism brings computational savings. While solving locally, large-scale Eigen value problem incurs $o(n^2)$ cost per-iteration, our proposed model incurs $o(n)$ computational burden for the client in each iteration [8]. From experimental result we have shown that as matrix dimension gets larger, client achieves better efficiency.

This paper is organized as follows: Section II shows theoretical background of Eigen value problem. Section III describes system model. In Section IV, we have represented our proposed model to solve the Eigen value problem. Theoretical analysis and experimental result is provided in Section V. Finally, we have concluded the work in section VI.

## II. THEORITICAL BACKGROUND

### A. Power Iterative Method

Boundary value problems, such as study of vibrating systems, structure analysis and electric circuit system analysis reduce to a system of equations of the form $Ax = \lambda x$. Such problems are called Eigen value problems, where $\lambda$ is a scalar constant which is called Eigen value and $x$ is Eigen vector. Due to easy implementation and less computational power consumption of power iterative method, many applications use it frequently. However, it will find only the largest Eigen value and corresponding Eigen vector. The power iteration algorithm starts with a random vector or an approximation to the dominant eigenvector $x^0$. The iterative equation can be represented as:

$$x^{k+1} = \frac{A.x^k}{\|A.x^k\|} \quad (1)$$

So, at every iteration, the vector $x^k$ is multiplied by an $n \times n$ (nonsingular) coefficient matrix $A$, and $\|A.x^k\|$ parameter is the scaling factor, which is the largest magnitude of $A.x^k$. When the iteration converges, we get the largest Eigen value and corresponding Eigen vector. This method can be more efficient for some computational tasks, although it finds only the largest Eigen value. For example, Google uses it to calculate the PageRank [3] of documents in their search engine.

### B. Homomorphic Encryption

Homomorphic encryption is a form of encryption that allows computations to be carried out on cipher text, thus generating an encrypted result which, when decrypted, matches the result of operations performed on the plaintext [9]. We used paillier cryptosystem [10], which is a partially homomorphic encryption system and has additive homomorphic property providing better efficiency than fully homomorphic scheme [11]. It is a public key cryptography that uses asymmetric key algorithm.

If two cipher texts are multiplied, encrypted form of addition of the plaintexts is formed. Decrypting this we get the sum of two plaintexts.

$$D(E(x_1, r_1) * E(x_2, r_2) \bmod n^2) = x_1 + x_2 \bmod n \quad (2)$$

If a cipher text and a plaintext raising $g$ are multiplied, encrypted form of addition of the plaintexts is formed. Decrypting this we get the sum of two plaintexts.

$$D(E(x_1, r_1) * g^{x_2} \bmod n^2) = x_1 + x_2 \bmod n \quad (3)$$

Practically, paillier cryptosystem holds following identities [12]:

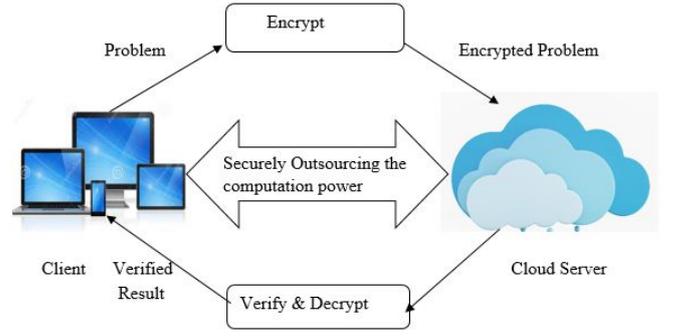

Fig. 1. Secure Outsourcing System Model [13]

$$D(E(x_1) * E(x_2) \bmod n^2) = x_1 + x_2 \bmod n \quad (4)$$

$$D(E(x)^k \bmod n^2) = kx \bmod n \quad (5)$$

$$D(E(x_1) * g^{x_2} \bmod n^2) = x_1 + x_2 \bmod n \quad (6)$$

$$D(E(x_1)^{x_2} \bmod n^2) = x_1 x_2 \bmod n \quad (7)$$

## III. SYSTEM MODEL

Fig. 1 illustrates the system model for securely outsourcing Eigen value problem. Client with low computational power outsources the problem to a cloud server having huge computational power and special software. To protect input privacy, the client first encrypts the original problem using a secret key $K$ to get encrypted problem. Then, the encrypted problem is given to the cloud for a result. Once the cloud receives encrypted problem, the computation is carried out with software. Then cloud sends back the result to client. After getting the returned result, the client then decrypts the result using the secret key $K$ and get original result. Then, the client checks the result whether it is correct or not. If correct, client accepts it, otherwise rejects it.

## IV. PROPOSED METHOD

As solving Eigen value problem by using direct method is time consuming for large matrix, it might not be a good option for resource-limited clients. Moreover, numerical techniques for approximating roots of polynomial equations of high degree are sensitive to rounding errors [14]. For these reasons we have used power iterative method that only finds the eigenvalue that is largest in absolute value and corresponding Eigen vector.

We have made following assumptions: The coefficient matrix $A$ has a dominant eigenvalue with corresponding dominant eigenvectors [14]. That means within all eigenvalues of $A$, one eigenvalue is strictly greater than the other eigenvalues in magnitude. The initial approximation of the starting vector $x^0$ must be a nonzero vector in $R^n$.

Our proposed protocol has three phases, Problem Transformation, Problem Solving and Result Verification.

## A. Problem Transformation

The iterative equation can be represented as:

$$x^{k+1} = \frac{A.x^k}{\|A.x^k\|} \quad (8)$$

In this equation, the most expensive calculation is matrix vector multiplication, $A.x^k$. It takes $o(n^2)$ time complexity. For this purpose, this costly calculation is carried out by cloud. But as cloud server can be malicious, client cannot send co-efficient matrix $A$ and $x$ directly to the cloud. For maintaining the confidentiality of data, client encrypts $A$ by paillier encryption, thus gets encrypted $A$ and then sends $Encrypted(A)$ to the cloud. Then in each iteration, client masks $x$ by a random vector $r \in R^n$ and gets $z$.

Where,

$$z^k = x^k + r \quad (9)$$

Then client sends $z^k$ to the cloud. Cloud performs operation on $z^k$ and returns encrypted version of $A.z^k$ to the client. Client decrypts and get,

$$A.z^k = A(x^k + r) \quad (10)$$

From (10), for getting $A.x^k$, client subtracts $c$ from $A.z^k$. Where, $c = Ar$. So,

$$A.x^k = A.z^k - c \quad (11)$$

Then client updates $x$ by (8) and then again mask $x$ by (9) for the next iteration and sends again to the cloud. This process continues until convergence.

## B. Problem Solving

In problem solving phase we have considered the overall steps for solving Eigen value problem. Cloud computes matrix-vector multiplication $A.z^k$ in each iteration. It is enough to describe the very first round of the iterative process to describe the whole problem solving phase.

1. At starting, client guesses the initial approximation of the eigenvector $x^0$, $x^0 = x_1^0, x_2^0 \ldots \ldots x_n^0$. Then client takes a random vector $r$ and forms $z^0$ by $z^0 = x^0 + r$. Client encrypts $A$ by paillier encryption and sends $Encrypted(A)$ and $z^0$ to cloud. Then generates $c$ by $c = Ar$.

2. On the encrypted matrix $Encrypted(A)$ and $z^0$, The cloud computes the value $Encrypted(A.z^0)$ by using the additive homomorphic property of paillier encryption:

$$Encrypted(A.z^0)[i] = Encrypted(\sum_{j=1}^{n} A[i,j].z_j^0)$$

$$= \prod_{j=1}^{n} Encrypted(A[i,j])^{z_j^0} \quad (12)$$

For $i = 1,2\ldots n$, and sends $Encrypted(A.z^0)$ to client [13].

3. The client decrypts $Encrypted(A.z^0)$ by using private key when it is received from the cloud and gets $A.z^0$. Client then gets $A.x^0$ using (11) and then updates the next approximation of $x$ using (8).

$$x^1 = \frac{A.x^0}{\|A.x^0\|} \quad (13)$$

This process continues until convergence occurs. Client checks for convergence in each iteration. In $k$ th iteration, client sends $k$ th approximation $z^k$ to the cloud and cloud calculates $Encrypted(A.z^k)$ and sends back to the client. Then client updates next approximation $z^{k+1}$ and sends again to the cloud. Client sends $Encrypted(A)$ to cloud only in the first iteration.

## C. Result Verification

Cloud server can be unfaithful and give the incorrect result to the client. So client needs to verify the result for detecting cheating. When iteration converges, client performs result verification by the following equation,

$$Ax = \lambda x \quad (14)$$

Where $\lambda$ the maximum Eigen value and $x$ is the corresponding Eigen vector. If result of $Ax$ becomes equal to $\lambda x$, client takes the result as a correct result. Otherwise rejects the result.

A truly malicious cloud server can return arbitrary answers in each iteration. As a result, a situation can occur when the iteration would never converge, wasting the resources of the client [13]. To prevent such wastage of resources, client can preset a sufficiently large value $\omega$ so that program will automatically converge after $\omega$ iterations. After convergence, client uses (14) to check the correctness of the result. Again, client can use multiple cloud providers to mitigate the risks associated with the malicious cloud.

## V. THEORITICAL ANALYSIS & EXPERIMENTAL RESULT

### A. Theoritical Analysis

*1) Convergence analysis:* When dealing with iterative methods, it is must to determine whether and when the iteration will converge [13]. Client defines an error limit $\in$, and determines convergence if,

$$\left\| x^k - x^{k+1} \right\| \leq \in \quad (15)$$

Where value of $\in$ is very small, close to zero.

*2) Input Security Analysis:* We can say that our proposed protocol protects client's input data privacy if the cloud cannot get the original matrix $A$ from the encrypted matrix $Encrypted(A)$ and original vector $x$ from the encrypted vector $z$ in each iteration [5]. We encrypt $A$ by paillier cryptosystem so cloud server cannot recover $A$ from $Encrypted(A)$. In each iteration of our proposed method, the cloud only sees the plaintext of $z^k$ and the encrypted version of matrix $Encrypted(A)$. Though we send $z$ as plaintext to the cloud, it is safe because original vector $x$ is randomly masked by a vector $r$. No information of $x$ would be leaked as long as $r$ is kept secret by the client [13]. For high security, each element of random vector $r$ should be 128 bits or more. Currently, if a supercomputer could check a billion billion ($10^{18}$) keys per second, then breaking a 128-bit key space by brute force requires about $3 \times 10^{12}$ years [15]. For enhancing the privacy of input data, we can further use a random scaling factor $a_k \in Z$ for each iteration to break the linkability of two consecutive iterations of the protocol [13]. Specifically, instead of sending $z^k$ to the cloud server, the client can send $a_k.z^k$ for the $k$ th iteration. Thus, the cloud server can not establish linear equations from received $a_k.z^k$ and $a_{k+1}.z^{k+1}$ [13].

*3) Output Security Analysis:* We can say that our proposed protocol protects output data privacy if the cloud cannot get the correct $A.x^k$ from $Encrypted(A.z^k)$ in each iteration. If the customer sends $a_k.z^k$ for the $k$ th iteration to the cloud, the cloud server sends back the encrypted result $Encrypted(a_k.A.z^k)$. Then the client just simply decrypts the $Encrypted(a_k.A.z^k)$ to get $a_k.A.z^k$ and divides each component with $a_k$. For the next iteration another random scaling factor $a_{k+1}$ is multiplied to $z^{k+1}$ before sent to the cloud server [13].

*4) Efficiency Analysis:* If client wants to solve the whole problem locally, client needs to compute matrix-vector multiplication $A.x^k$ in each iteration, which takes $o(n^2)$ complexity. This costly computation in each iteration incurs huge burden for client. On the other hand, in our proposed protocol, in the initial setup phase, client needs to encrypt $A$ by paillier cryptosystem which takes $o(n^2)$ complexity and generate c by $c = Ar$ that takes $o(n^2)$ complexity. Client needs to perform these two $o(n^2)$ computations only once. Then in all the iterations client only needs to form $z^k$, decrypt the vector of $Encrypted(A.z^k)$ to get $A.z^k$, generate $A.x^k$ from $A.z^k$ and update next approximation. Each of these operations take only $o(n)$ complexity. After convergence, result verification is performed by the client that takes $o(n^2)$ complexity. Expensive matrix-vector multiplication in each iteration is carried out by the cloud instead. As a result computational burden decreases for the client and efficiency increases. Again, as the encryption on each element of the coefficient matrix $A$ is independent, it can be easily parallelized [13]. For this reason, it is not mandatory for the client to load the whole coefficient matrix in memory in the first place and client's computational burden further decreases.

### B. Experimental Result

In our experiment, we have used the same machine both for cloud and client. Then we have analyzed the performance of the client side. We have observed how much efficiency is achieved by the client by using our proposed protocol and how much computational burden is reduced due to outsourcing. We have ignored the communication latency between the client and the cloud for this application since the computation dominates the running time as shown in our experiment [5]. The experiment is carried out using Eclipse on a workstation equipped with Intel(R) Core(TM) i3-2350M CPU @2.30GHz and 4GB RAM. In our experiment, we have implemented power iterative method to solve Eigen value problem where the most expensive calculation matrix-vector multiplication is employed by the cloud.

Our main goal is to provide efficiency to the client by using outsourcing. For this purpose, efficiency gain is calculated. Efficiency gain can be termed as ***Client Speedup***. Here, ***Client Speedup*** is the ratio of the time that is needed for client if the computation is done locally and the time that is needed by the client's computation if outsourcing is chosen [5]. Definition of some parameters used in our analysis are provided in Table I.

TABLE I. DEFINITION OF NOTATIONS [5]

| Notation | Meaning |
| --- | --- |
| $t_{original}$ | Required time for the client to solve the original problem locally if outsourcing is not used. |
| $t_{cloud}$ | Required time for the cloud to compute the calculations that are sent to the cloud. |
| $t_{client}$ | Required time for the client to perform encryption, decryption and result verification if outsourcing is used. |
| Client Speedup | $\dfrac{t_{original}}{t_{client}}$ |

TABLE II. EXPERIMENTAL RESULT FOR EIGEN VALUE PROBLEM

| Data Dimension, $n$ | Original problem $t_{original}$ | Encrypted Problem $t_{cloud}$ | Encrypted Problem $t_{client}$ | Client Speedup $\dfrac{t_{original}}{t_{client}}$ |
|---|---|---|---|---|
| 50 | 0.3133 | 0.3130 | 0.1903 | 1.6463 |
| 100 | 1.2921 | 1.2911 | 0.6037 | 2.1403 |
| 200 | 7.3012 | 7.3011 | 2.2261 | 3.2798 |
| 300 | 21.6815 | 21.6814 | 4.9284 | 4.3993 |
| 400 | 47.5771 | 47.5761 | 9.1275 | 5.2125 |
| 500 | 84.9826 | 84.9820 | 13.3055 | 6.3870 |
| 750 | 273.7168 | 273.7161 | 31.7513 | 8.6206 |
| 1000 | 485.7088 | 485.7082 | 51.0005 | 9.5236 |

Our experimental result is given in Table II. We can observe that client speedup increases as matrix dimension $n$ gets larger. In each iteration, $o(n^2)$ computation is performed by cloud and $o(n)$ computation is performed by client. As a result client gets desired computational savings when $n$ goes sufficiently large. We have also noticed that the value of **Client Speedup** is greater than 1 that means we have achieved desired performance gain.

As we have implemented the experiment in same workstation, that means we have considered client's machine as both client and cloud, we have got the time required for cloud is sufficiently large. Time required for cloud is given in third column. As the real cloud is equipped with high performance parallel computing and virtualization, matrix-vector multiplication can be divided into subtasks either by Block-striped Matrix Partitioning or by Checkerboard Block Matrix Partitioning [16] [17]. As a result, if real cloud is used, time required for cloud to perform matrix-vector multiplications would be much smaller comparing with the result achieved in our experiment as we have considered client's machine as cloud. In that case, we would be able to modify the equation for **Client Speedup** as,

$$Client\_Speedup = \frac{t_{original}}{t_{client} + t_{cloud}} \quad (16)$$

In case of real cloud, if we use (16) to calculate **Client Speedup**, we can assume that we would still achieve significant speedup as $t_{cloud}$ would be very small for real cloud.

V. CONCLUSION

In this paper, we have proposed a protocol for solving Eigen value problem using outsourcing. Concerning about client's burden in case of solving large-scale Eigen value problem, we have proposed an idea of outsourcing that would increase the client's efficiency and lowers the burden of client. In parallel with providing efficiency to client, this protocol also provides security to client's confidential data and result verification. This protocol reduces client's computational burden and provides better efficiency. Directions for further research include: 1) Implementing this protocol using real cloud. 2) Identifying a better result verification technique that would perform result verification in each iteration.3) Identifying new meaningful scientific and engineering computational tasks and then designing protocols to solve them [5]. 4) Adding result verification for some existing protocols that do not have result verification mechanism